\documentclass{iopart}

\usepackage{graphicx}

\usepackage[usenames,dvipsnames]{color}
\usepackage{mathptmx}
\begin{document}

\title{Hollow Microspheres as Targets for Staged Laser-Driven Proton Acceleration}
\author{M Burza$^1$, A Gonoskov$^{2,3}$, G Genoud$^1$, A Persson$^1$, \\ K Svensson$^1$, M Quinn$^4$, P McKenna$^4$, M Marklund$^2$, \\ C-G Wahlstr\"om$^1$}

\address{$^1$ Department of Physics, Lund University, P.O. Box 118, SE-221 00 Lund, Sweden}
\address{$^2$ Department of Physics, Ume{\aa} University, SE-901 87 Ume\aa, Sweden}
\address{$^3$ Institute of Applied Physics, Russian Academy of Sciences, 46 Ulyanov Street, Nizhny Novgorod 603950, Russia}
\address{$^4$ SUPA Department of Physics, University of Strathclyde, Glasgow, G4 
0NG, United Kingdom}
\ead{matthias.burza@fysik.lth.se}

\begin{abstract}
A coated hollow core microsphere is introduced as novel target in ultra intense laser matter interaction experiments. In particular, it facilitates staged laser driven proton acceleration by combining conventional target normal sheath acceleration (TNSA), power recycling of hot laterally spreading electrons and staging in a very simple and cheap target geometry. {During TNSA of protons from one area of the sphere surface, laterally spreading hot electrons forms a charge wave. Due to the spherical geometry, this wave refocuses on the opposite side of the sphere, where an opening has been laser micromachined. This leads to a strong transient charge separation field being set up there, which can post-accelerate those TNSA protons passing through the hole at the right time.} Experimentally, the feasibility of using such targets are demonstrated. A redistribution is encountered in the experimental proton energy spectra, as predicted by particle-in-cell simulations and attributed to transient fields set up by oscillating currents on the sphere surface.

\end{abstract}

\maketitle

\section{Introduction}

Laser-driven ion acceleration is an area of research that currently attracts significant scientific interest. The ion beams produced in these experiments have several attractive characteristics, such as very low transverse emittance and small virtual source size \cite{Cowan} together with a short pulse duration (at the source). Proposed applications of this possibly compact ion beam source include ion radiotherapy for cancer treatment \cite{Malka, Linz}, isotope production for medical imaging techniques \cite{Spencer}, proton radiography of inertial fusion plasmas \cite{Mackinnon} and implementation as injectors for future ion accelerators. 

In a typical experiment, a high power laser pulse of short duration, $\leq$\,ps, is focused on the surface of a thin foil to an intensity exceeding  $10^{19}$\,W/cm$^2$. The laser interacts with target electrons and a population of hot electrons with a Maxwellian temperature of typically a few MeV is generated. A large fraction of these electrons traverse the target and build up exceptionally high electrostatic fields, $\sim$ TV/m, at the rear surface of the foil, in a direction normal to the target surface. Atoms on the target surface are rapidly field-ionized and accelerated. This is referred to as Target Normal Sheath Acceleration (TNSA) \cite{Wilks}. Because of the presence of hydrocarbon and water vapour on the surfaces of the foils (at typical vacuum conditions $\sim10^{-5}$ mbar), protons are the dominating ion species. Due to their high charge-to-mass ratio, protons are more efficiently accelerated than heavier ions.  

The acceleration of protons behind the target foil is very rapid, due to the high field strength. However, this field is present during a short time only, limiting the maximum energy reached by the protons. The energy spectra of these proton beams exhibit a longitudinal emittance {comparable to that of conventional accelerators}, with a quasi-exponential shape and a distinct cut-off energy \cite{Wilks}. The divergence of the proton beam is typically $\sim30^\circ$ half angle. Significant theoretical and experimental efforts have been devoted to the exploration of means to boost the maximum proton energy without the use of increasingly larger laser systems \cite{Glinec, McKenna-LPB-2008}. 

Practical limitations in laser size and costs, laser materials and repetition rate are craving for alternative or modified laser acceleration schemes and targets to further increase the peak proton energy. It has been found {that the maximum proton energy and laser-to-ion energy conversion is enhanced by the use of ultra thin targets in combination with laser pulses of high temporal contrast \cite{Neely}. Staging, i.e. combining two or more accelerator stages in series, may be a way to post-accelerate a selected portion of the protons accelerated in a preceding TNSA stage and thus raise the maximum proton  energy and  reduce the energy spread \cite{Jaeckel}. In parallel, extensive studies on controlling beam parameters such as collimation and means to produce quasi-monoenergetic energy distributions have been carried out \cite{Allen, Brantov}. In particular, mass-limited targets can be used to reduce the energy spread of the protons \cite{Schwoerer, Sokollik,Buffechoux}. Curved target foils \cite{Roth1}, electrostatic charging of specially shaped targets \cite{Kar} or separate focusing cylinders \cite{Toncian}, enable spatial shaping of the proton beam.}

 In addition, experiments and numerical modelling have shown that while part of the hot electron distribution is passing through the target foil, a significant part is spreading also laterally along the target. McKenna \textit{et al.} \cite{McKenna} found that, when these electrons reach the target edges, after a time determined by the geometrical size of the target and the lateral electron transport velocity, they  establish quasi static electric fields, similar to the one produced behind the target during TNSA, resulting in ion acceleration from the edges. Normally, this mechanism just represents a loss of absorbed laser energy, which is converted to hot electrons but not contributing to the quasi static sheath built up at the target rear side. In a recent study \cite{Buffechoux}, however, using very small diameter targets, the refluxing of transversely spreading electrons were found to enhance and smooth the sheath field for TNSA from the rear surface.

In this paper we discuss the use of hollow microspheres, as novel targets for laser acceleration of protons. With this target, several of the above features are combined, which may facilitate improved laser-to-proton efficiency, {eventually leading to} increased proton energy and reduced divergence. Lateral electron transport is here utilized to set up a post-acceleration field for staged acceleration.\\

\begin{figure}[htbp]
\centering
\includegraphics[width=7cm]{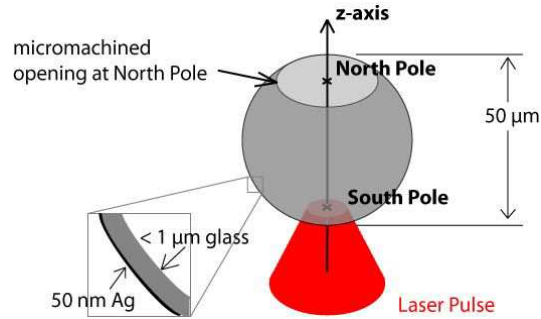}
\caption{A glass hollow microsphere,  with an $\sim 50$\,nm silver coating on its $\leq  1 \mu$m thick wall is struck by a laser pulse at the "south pole". TNSA protons are emitted though a circular opening at the "north pole" and post-accelerated}
\label{fig:spheregeometry}
\end{figure}

The basic idea behind our approach is to use hollow microspheres with diameters of about $10-50$\,$\mu$m and sub-micrometer wall thickness. In each sphere a small circular opening is made. We refer to the position of this opening as the "north pole" (see figure \ref{fig:spheregeometry}). A short pulse laser irradiates the sphere at the "south pole", where TNSA  takes place. The primary proton direction will be along the z-axis, defined as the axis from the south pole passing through the north pole of the sphere. The spherical surface, with TNSA taking place from the concave side, results in a collimated or even converging proton beam. Therefore, all the protons can be made to pass through the opening at the north pole. In addition -  and this is the key point - electrons  leaving the laser focus laterally in any direction along the sphere surface, will be guided on different longitudes over the sphere and eventually reach the edge of the opening at the north pole simultaneously after some given time. A very strong quasi static electric field is then formed in the opening, along the z-axis.  {This quasi-static field will post-accelerate protons passing through the opening at the correct time. }

{In our approach to test this idea, theoretical and experimental studies go hand in hand:}

{To test the experimental feasibility, we perform experiments, at the Lund High Power Laser Facility, with commercially available hollow microspheres of $50$\,$\mu$m diameter$^1$.  The walls of these spheres are made of glass with a thickness of $0.5-1$\,$\mu$m, and coated with a $\sim 50$\,nm silver layer (see inset in figure \ref{fig:spheregeometry}), which facilitates optical alignment and guiding of electrons along the sphere surface. Openings of different sizes are laser micromachined.} 
\footnotetext[1]{Such spheres are very low weight and low cost objects. They weigh only some tens of ng each and cost $\sim 1$\,USD for $10^5$ {Ag coated spheres. The manufacturing of the hole at the north pole, however, is part of the local target preparation and not included in the price.}}
{We present these experiments in Section 2, including target preparation, fixation and alignment in the experimental setup together with first results.}

{In parallel, we perform Particle-In-Cell (PIC) simulations of hollow conducting spheres with openings,  irradiated by short laser pulses. These simulations, presented in Section 3, qualitatively describe the dynamics involved.} 



We discuss the outlook and prospect for further experiments in Section 4, and conclude in Section 5.

\section{Experiment}

\subsection{Target preparation}

Isolated spheres are suspended into a nylon mesh grid, where both the front and back side of each sphere is accessible for further processing.
Openings in the spheres are made with a confocal microscope based laser micromachining setup. In a two step process the silver coating is ablated in a well defined region on the sphere surface, followed by ablation of the glass substrate. This is done with a lateral resolution of about $2$\,$\mu$m utilizing femtosecond laser system running at $10$\,Hz repetition rate. Real time target observation and a high numerical aperture in the setup facilitate both, high drilling accuracy and control in the transverse direction while at the same time preventing the TNSA surface inside the sphere from being damaged in the procedure.

Afterwards the target, which is still fixed in the nylon mesh, is mounted into a holder. {This also accommodates a gold mesh, placed close to the sphere's north pole, to extract information about proton trajectories, as will be discussed later.}

\subsection{Experimental setup}

The Lund multi-terrawatt laser, which is a Ti:Sapphire system based on chirped pulse amplification (CPA), is used for this experiment. In this experiment, it is tuned to $850$\,mJ pulse energy at $805$\,nm centre wavelength with a typical pulse duration of $42$\,fs FWHM. Due to the sub micron thickness of the sphere walls, and the thin silver coating on them, an amplified spontaneous emission (ASE) contrast better than $10^{8}$ some tens of picoseconds prior to the pulse peak is desirable. 

\begin{figure}[htbp]
\centering
\includegraphics[width=7.5cm]{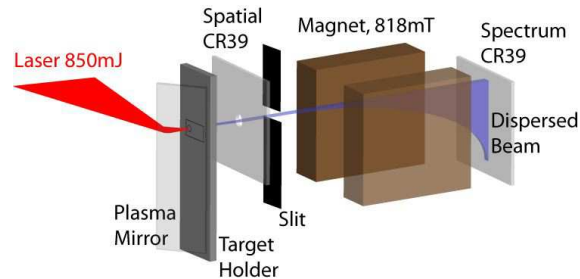}
\caption{The laser pulse (red) is impinging on the PM at Brewster's angle, which reflects the pulse normal onto the target. After some centimetres of free passage the resulting particle beam (blue) reaches a CR-39 detector plate, which has the function of providing a lateral image of the beam profile and at the same time enabling a fraction of the beam to enter slit of a subsequent {permanent} magnet spectrometer. After traversing the field, a vertically dispersed spectrum is recorded on a second CR-39 plate.}
\label{fig:expsetup}
\end{figure}

To optimize contrast on a very fast time scale, the convergent, horizontally polarized, laser pulse hits a dielectric plasma mirror (PM) at Brewster's angle $(3.0 \pm 0.2)$\,mm prior to the primary focus. At this location, the plasma mirror is operating at $(8.5 \pm 1.1) \times 10^{15}$\,W/cm$^2$ spatially averaged peak intensity over the beam diameter ($I_{\rm centre}/e^2$). When activated, it deflects the laser beam normal onto the target (see figure \ref{fig:expsetup}).
  
Plasma mirror characteristics have been investigated by many groups (see e.g. \cite{Dromey,Doumy,Ziener}). A PM assembly, similar to the one applied in the present experiment was utilized by Neely \textit{et al.} \cite{Neely}, using the Lund laser system. In that experiment proton beams from Al foil targets as thin as $20$\,nm were observed. Our experiment relates to that one as the very thin silver coating on our target surface is of comparable thickness. 
Ray tracing, taking a $22$\,nm FWHM broad Gaussian spectrum and a p-polarized converging beam into account, predicts a contrast increase by a factor $100$ in our experiment (assuming a maximum reflectivity $\sim 50$\,\% from the PM \cite{Ziener}).
  
Together with a $3^{rd}$ order autocorrelation contrast measurement, $2$\,ps prior to the main pulse, a contrast better than $10^{6}$ on target can be guaranteed for an intact rear TNSA surface during the first phase of acceleration. This contrast is due to non-perfect compression in the CPA chain and should not be mistaken for the ASE contrast, encountered on a longer picosecond timescale prior to the main pulse, which is of the order $10^{11}$ on target.
  
The infrared (IR) pulse is focused by an $f/3$ off-axis parabolic mirror (OAP) down to a $4.4$\,$\mu$m spot diameter (intensity FWHM), containing $39 \%$ of the total energy and reaching a peak intensity of $\sim 3 \times 10^{19}$\,W/cm$^2$.  Target positioning is accomplished by a confocal imaging system: An expanded HeNe laser beam is superimposed with the IR and a confocal reflection from the silver coated target surface is imaged, utilizing the OAP as an objective. 

The detector system for protons, designed to simultaneously provide a spatial beam profile and a spectrum, is depicted in figure \ref{fig:expsetup}. It consists of a primary CR-39 plate at some centimetres distance from the target, which is utilized to characterize the transverse spatial beam profile. This plate is covered by a $6$\,$\mu$m Al foil, which stops protons below $0.5$\,MeV \cite{Pstar}. In its centre, a $\sim 4$\,mm diameter hole enables protons close to the target normal axis to continue to the momentum dispersive part of the detector and access an $88$\,$\mu$m wide entrance slit of a permanent magnet spectrometer, where they traverse a $51$\,mm long and $818$\,mT strong effective field. The vertically dispersed proton spectrum is then recorded by a second CR-39 plate with an accuracy of $\pm 0.2$\,MeV at $4$\,MeV proton energy. By this arrangement spectra can be correlated to the lateral position within the particle beam.

\subsection{Experimental measurements and results}

After 
optimization of the plasma mirror working distance with the help of proton beams originating from flat $400$\,nm thick Al foil targets, shots were taken on machined and unmachined microspheres as well as on $0.9$\,$\mu$m thick Mylar foil targets.

\begin{figure}[htbp]
\centering
\includegraphics[width=6cm]{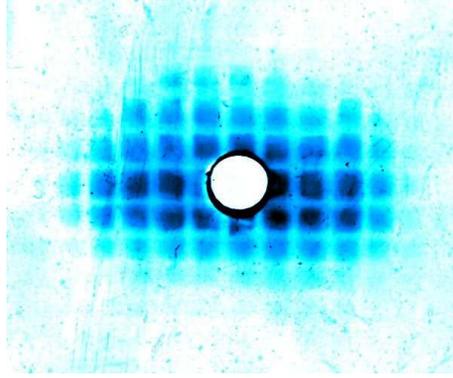}
\caption{Proton beam profile on a CR-39 plate showing a magnified Au mesh image, recorded $(24 \pm 0.5)$\,mm distance from the target}
\label{fig:cr39_1659}
\end{figure}

Several shots were taken on machined microspheres with holes of typically $18$\,$\mu$m diameter. An example of a proton beam imprint originating from such a sphere on the spatial CR-39 plate, covered by $6$\,$\mu$m Al and located at $(24.0 \pm 0.5)$\,mm distance from the target, can be seen in figure \ref{fig:cr39_1659}.

One can see a slightly oval beam profile. 
This slight asymmetry can be attributed to either a non uniform Ag coating on the sphere's south pole or to grazing incidence of the laser energy around the z-axis due to the curved sphere surface: At a given latitude, the linear polarization of the laser pulse will be incident as p or s polarization on the spherical surface, depending on the azimuth angle. Polarization effects will also facilitate different strengths of the surface currents, depending on the longitude where electrons propagate around the sphere \cite{Psikal}.

One can further see the imprint of the earlier mentioned Au mesh on the spatial proton beam image, which is introduced into the proton beam path close to the target. This rectangularly shaped $4$\,$\mu$m thick mesh features squared holes with a nominal aperture of $11 \times 11$\,$\mu$m$^2$ and a lateral wire widths of $5$\,$\mu$m. It is fixed at $(165 \pm 2)$\,$\mu$m distance from the rim of the opening of the microsphere. {The motivation to introduce this mesh is twofold: Firstly, it verifies, that protons contributing to the signal on the spatial CR-39 plate are not due to edge emission at the north pole, which would have resulted in a distinctively different shadow image. Secondly, thanks to this mesh the protons can be shown to come from a virtual source located $(54 \pm 12)$\,$\mu$m from the inner sphere TNSA surface very close to the opening. This estimate is valid for the majority of particles traversing the opening at late times, where we expect the surface oscillations to have vanished. In order to visualise effects of a transient field emanating from the rim of the microsphere opening, one would have to filter this image to the appropriate energy. However, as will be discussed later, with our present experimental parameters, we expect only the very fastest particles to be affected by these fields, so a filtered signal would become very weak.}

{By neglecting Coulomb repulsion between protons in the beam, and tracing proton trajectories further back, one can make a rough estimate about the proton emission surface. We find that proton emission seems to occur from a solid angle} covering $\approx 140 \pi$\,msr of the inner shell, measured from the centre of the sphere. This compares to a focal spot, covering $\approx 8 \pi$\,msr (intensity FWHM) and is consistent with previous observations of the TNSA surface source area being considerably larger than the laser focus {\cite{Roth3, Schreiber}}. 
The virtual proton source in a flat foil TNSA experiment \cite{Borghesi1} was pinpointed to several hundred microns distance before the target front side, i.e. the laser irradiated side. A spherical target, such as the one used for isochoric heating \cite{Roth2, Patel}, combines ballistic proton propagation with target curvature and an altered electron distribution. A particle focus near the north pole is consistent with these findings.

To further verify that the protons are indeed emitted from the sphere interior, we {irradiate} closed hollow microspheres that have not gone through the laser micromachining stage. {Lacking hydrogenic contaminations, such as water vapour, on the interior surface of the sphere, TNSA of protons is not expected to occur there. Indeed no protons with energies sufficient to be recorded by our diagnostics ($E_{proton} > 0.8 $\,MeV) are observed. In addition, lacking the opening and retaining the silver coating, return currents will prevent the formation of a strong edge field at the north pole. This will be further discussed in section 3.1.}
 
\begin{figure}[htbp]
\centering
\includegraphics[width=14cm]{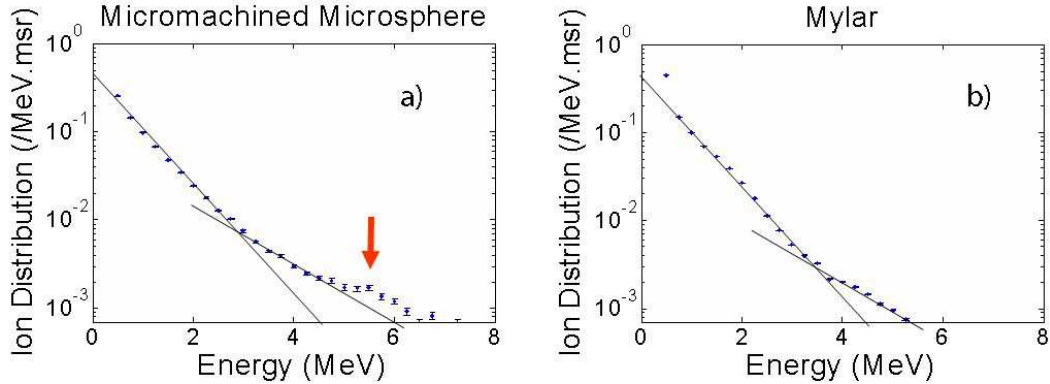}
\caption{{Normalized proton spectra from a Microsphere target (a)) and a Mylar plane foil (b)); The red arrow in a) indicates a region of constant yield in contrast to the strictly decreasing number density from plane foil TNSA experiments, such as the one depicted in b); black lines are a dual temperature guide for the eye}}
\label{fig:spectracompilation}
\end{figure}

\nopagebreak[4]
Spectra from sphere targets with openings between $18$ and $20$\,$\mu$m where taken and a typcial spectrum can be seen in figure \ref{fig:spectracompilation} a). A clear high energy "cut off" is not visible in the data, and values above $8$\,MeV are ignored to assure that the data presented here are at least one order of magnitude above the noise level.
Reference shots are taken on flat $0.9$,$\mu$m thick Mylar foil targets. A typical proton spectrum can be seen in figure \ref{fig:spectracompilation} b). Even though the laser absorption and particle yield is expected to be different as compared to the silver coated glass surface of the microspheres {($\sim 30$  times higher particle number)}, those shots provide reference spectra, enabling the identification of special features  in the microsphere spectra.

All microsphere spectra from the experimental study look very similar, {but with an integrated particle yield varying by a factor 4, which is twice as much as for the Mylar targets}. There are indeed features present, that could possibly be attributed to a secondary field interaction and a post-acceleration by a secondary field at the sphere opening, which is still prevailing {during arrival of the fastest protons}. In all microsphere spectra there is a slight modification of particle yield between $5.5$\,to\,$6.5$\,MeV, where the counts per energy bin remain nearly constant (Figure \ref{fig:spectracompilation} a), red arrow) as compared to the strictly decreasing dual exponential decay observed from the flat foil targets, {shot under the same experimental conditions} (Figure \ref{fig:spectracompilation} b)). {(Likewise, previous experiments carried out with various planar targets using the Lund laser system \cite{McKenna2006} have resulted in spectra very similar to the ones presented here for Mylar foils.) The plateau in the microsphere spectra can be understood in terms of a spectral redistribution of the order $1$\,MeV of proton energies as a result of a secondary field interaction near the opening.} Looking at the experiment the other way around, we can regard the protons as a probe for intra cavity fields as well. {A two-temperature curve fit provides comparable temperatures for low energies for both targets ($(0.6 \pm 0.1)$\,MeV and $(0.5 \pm 0.3)$\,MeV for spheres and Mylar, respectively) but larger values for the high energy component from the microsphere spectra ($(2.0 \pm 0.1)$\,MeV) as compared to the foil spectra ($(1.6 \pm 0.3)$\,MeV).}\\
\nopagebreak[4]

In the following section we will discuss modelling of the present experiment, giving insight into the dynamic processes of hot electron transport on the sphere surface and {motivate the observed spectral features}. Beyond that, we will {consider} means to enhance the post-acceleration mechanism.

\section{Simulations}

To improve our understanding of the underlying dynamic processes, we have carried out a number of numerical experiments with the parallel PIC code Extreme Laser Matter Interaction Simulator (ELMIS), developed by the SimLight group \cite{SimLight}. ELMIS is a relativistic code, which uses a parallel Fast Fourier Transform (FFT) technique to solve Maxwell's equations.

The processes involved in the setup are essentially of two-dimensional (2D) nature. Thus, we perform 2D simulations in order to retain the appropriate space and time scales, while still not compromising the physical outcome. {However the 2D nature restricts the interpretation of these results to a qualitative level as scaling laws behave differently in two dimensions as compared to three dimensional space.}

In the simulations a linearly polarized TEM$00$ laser pulse (where electric field lies in the plane of simulation) with $\tau_l =50$\,fs duration (Gaussian profile, FWHM) and a total energy of $1$\,J is focused to a $10$\,$\mu$m spot on the target surface. The laser field reaches a maximum field strength equal to $\sim 3.5 \, a_{\rm rel}$, where $a_{\rm rel} = 2 \pi m c^2/(e \lambda) \approx 3.2 \times 10^{12}\, \mathrm{V/m}$, $e$ and $m$ are the electron charge and mass, respectively, $c$ the speed of light and $\lambda = 1 \, \mu\mathrm{m}$ is the laser wavelength. This corresponds to a maximum intensity of $2 \times 10^{19}$\,W/cm$^{2}$. The target consists of a hollow metal sphere with $D_S = 32$\,$\mu$m diameter and a $10$\,$\mu$m opening. It has a $0.5$\,$\mu$m thick wall, which in our 2D PIC simulation was considered as a (cylindrical) overdense plasma with the electron density $50$\,$n_{\rm crit}$ and an Au$^{6+}$ ion density of $50$\,$n_{\rm crit}/6$, where $n_{\rm crit} = \pi m c^2/(\lambda^2 e^2) \approx 1.1 \times 10^{21} \, \mathrm{cm}^{-3}$ is the critical density for $\lambda = 1$\,$\mu$m.
 
To simulate TNSA, we consider a $100$\,nm contaminant layer of protons and electrons with a density $10 \, n_{\rm crit}$, covering the internal surface of the target. The simulation is done for a box size of $64$\,$\mu$m\,$\times$\,$64$\,$\mu$m ($4096 \times 4096$ cells) with absorbing boundaries for the fields and accumulating boundaries for the particles. The initial plasma temperature is set to 16 keV, and the cell size is 15.625 nm, which is approximately 4 times the Debye length for the considered plasma. In the simulation 100 virtual particles per cell are used for the electrons and Au$^{6+}$ ions and 20 particles per cell are used for the protons; the total number of virtual particles is $4 \times 10^7$. The time steps are set to $(2\pi/\omega_p)/16 \approx 3 \times 10^{-17} \, \mathrm{s}$, where $\omega_p = (4 \pi e^2 50 n_{\rm crit}/m)^{1/2}$ is the plasma frequency.
The duration limit of the simulation is set to when the leading protons in the accelerating bunch reach the simulation box boundary. 

\subsection{Simulation results} 

As the laser pulse reaches the target at the south pole it is reflected from the outer overdense plasma surface, initiating electron heating. This time is set to $t=0$\, in the simulation. Subsequently protons undergo TNSA from the internal surface and move towards the north pole. A part of the heated electrons leave the plasma, thereby producing an electric field retaining part of the electrons to the target surface.  Those trapped hot electrons move along the plasma layer, {recirculating near the wall and} conserving their momentum in the direction along the surface. Eventually they will arrive at the edge of the sphere opening, where they will leave and return to the plasma layer, thus setting up a charge separation field. This process, albeit for a flat target, was discussed and experimentally observed by McKenna \textit{et al.} \cite{McKenna}. Due to the relativistic intensity of the laser pulse, the electrons move with a speed close to $c$, thus forming a bunch size comparable to the longitudinal extension of the laser pulse ($c \tau_l = 15\,\mu$m). This bunch does effectively not carry any charges due to {cold return currents within the plasma. However, due to the absence of a return current at the edges, the bunch produces a charge separation field.}
The simulation shows that, this wave then reflects from the opening at the north pole and heads back to the south pole, where it refocuses, passes through (collisionlessly) and continues moving towards the opening at the north pole again. 

\begin{figure}
\centering
\includegraphics[width=\textwidth]{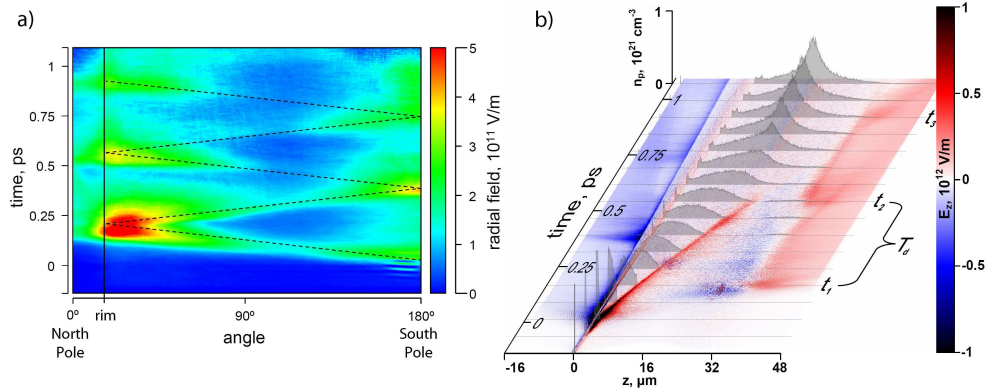}
\caption{{a) Radial electric field strength, encountered at $3\,\mu$m distance from the surface as function of time b) Electric field component $E_z$ along the z-axis; grey plots depict the proton density evolution along the z-axis in equidistant frames}}
\label{fig5}
\end{figure}

{The dynamics are illustrated in figure \ref{fig5} a), where the radial electric field at $3\,\mu$m distance from the sphere surface is plotted over time and latitude angle ($180\,^\circ$ corresponds to the south pole while $0\,^\circ$ represents the north pole). Figure \ref{fig5} b) depicts the electric field component $E_z$ along the z-axis, using red and blue colours. The evolution of the proton density along the $z$-axis is visualized by the grey distributions, in equidistant frames. In both pictures, one can identify large charge separation fields, set up at the rim due to the absence of a return current.} 

One can identify certain frames where the field becomes large, alternatingly at the north and south pole. The periodicity can be estimated as $T_d \approx \pi D_S/v_e$, where $v_e$ is the velocity of the surface dipole wave. From our simulations we obtain $T_d \approx 380$\,fs, corresponding to a frequency of 2.64 THz and an electron wave velocity of $v_e \approx 0.9\,c$.

\begin{figure}
\centering
\includegraphics[width=7.5cm]{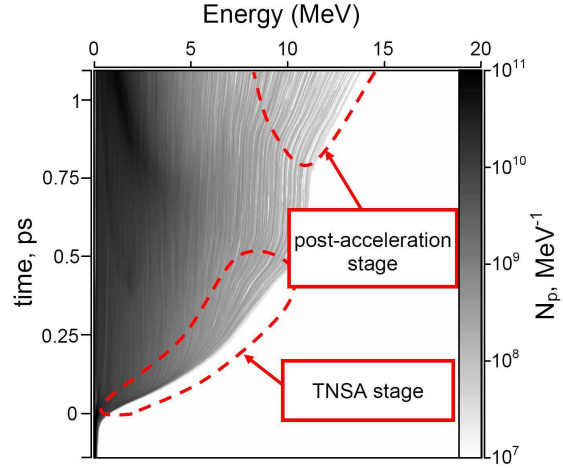}
\caption{spatially integrated proton energy distribution along the z-axis with color coded number density and its evolution in time}
\label{fig6}
\end{figure}

When the surface dipole wave reaches the opening it produces electric field maxima at the north pole at times $t_1$, $t_2$, $t_3$ (see labels in figure \ref{fig5} b)), which can be utilized for proton post-acceleration, i.e. staging. Those maxima exist only during relatively short periods. Thus, the timing of the proton bunch and the electron dipole wave is important. In order to have an effective post-acceleration, one would like protons to be pre-accelerated by the TNSA mechanism in such a way that they traverse the vicinity of the north pole when a maximum in accelerating field is present, resulting in a redistribution of the proton energy spectrum. It should be noted that, the surface dipole wave oscillation amplitude decays which implies that fewer oscillations prior to the proton passage provide a stronger post-acceleration. Additionally to the maxima, a weaker quasi constant accelerating background field occurs at the opening from $t_1$ onwards, showing no significant decay on this time scale. Post-acceleration by this background field, {which occurs due to a charge up of the sphere during laser irradiation}, does not require an accurate timing for the proton passage. However, it provides smaller field strengths as compared to the surface dipole wave oscillations. {The effect from both contributions can be seen in figure \ref{fig6}} in the upper marked region, labelled "post-acceleration stage", where the evolution of the proton energy distribution is displayed as a function of time. (Note that, due to the limited number of particles, modulations in these spatially integrated spectral distributions manifest themselves as lines that could easily be misinterpreted as trajectories.)

\begin{figure}
\centering
\includegraphics[width=7.5cm]{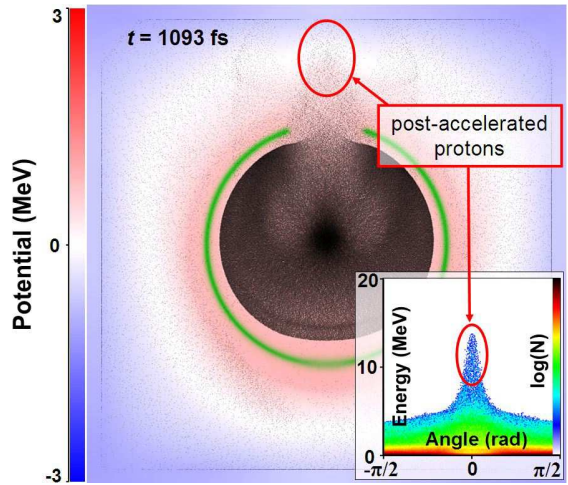}
\caption{The final instant of the simulation and the corresponding angle-energy distribution of the protons}
\label{fig7}
\end{figure}

The final state of this simulation is summarized in figure \ref{fig7} together with the {emission} angle-energy distribution in the  inset, taking all protons into account. From this figure it can be seen that, the considered geometry provides additional energy of several MeV for some part of protons which form a bunch with a small divergence of about $8^\circ$ half angle.

\section{Discussion and outlook}

In the present simulation, a fraction of the protons, with kinetic energies around $9$\,MeV, reach the opening at time $t_3$ and pass through a spike in accelerating field of the post-acceleration stage. However, the acceleration exerted on the protons by this field is relatively small. In 2D simulations the  charge increase due to refocusing of electron trajectories at the north pole is not fully reproduced. Even though qualitatively correct, the simulations are therefore expected to underestimate the field effects. 

However, a larger fraction of the particles could be post-accelerated if the relative timing between the surface dipole wave and the proton arrival could be controlled. This could be done by either making use of smaller spheres or oblate spheroidal targets to compensate for their different propagation velocities, or by simply increasing the proton temperature in the TNSA stage as they are still moving non-relativistically. In the latter case an energy of $9$\,MeV is sufficient for protons to reach the north pole at time $t_3$, while $37$\,MeV would be required for a passage at $t_2$ and GeV energies for $t_1$. 
TNSA acceleration to the GeV regime is not feasible, but $37$\,MeV should be within reach {of short pulse laser systems at intensities below $I_{laser} = 10^{21}$\,W/cm$^2$ \cite{Fuchs2006, Zeil2010}}. In addition, one might be able to reduce the velocity of the surface electron wave by surrounding the sphere with an appropriate dielectric. 

With spheroids the advantages of a spheres are conserved, but the relative distance for protons and electrons to propagate can be varied. Simulations with oblate spheroidal surfaces, performed as for the spherical targets above, show both stronger acceleration and a narrower collimation of the protons. Experimentally however, such targets are not as easily available as spheres.

In the experiment, we irradiated the target at normal incidence to obtain maximum symmetry and facilitate direct comparison with our simulations. However, it is well known that by irradiating the target with p-polarized light at an angle, {the efficiency of coupling laser energy to the plasma increases}. This should, in our case, {enhance} both TNSA at the south pole and drive a significantly stronger transverse electron current along the target surface \cite{Psikal}. An extension of both simulation and experimental geometry to allow for non-normal incidence irradiation will be a topic of further study. 

Finally, we would like to point out that our target has additional interesting features. One of them is an intermediate particle focus slightly outside the spatial boundaries of the target. This could be used for experiments in fundamental physics that require high proton flux, inherently synchronized with a high power laser beam line. Additionally, as the microsphere acts as a cavity for electron surface dipole waves, it could provide efficient means to produce THz radiation with high intensity lasers. In such an experiment almost all the incident laser energy can be absorbed by irradiating the sphere through the opening at the north pole, launching an electron surface wave from the inside.

\section{Conclusions}

{We have introduced a new scheme for staged laser-driven proton acceleration, using hollow microspheres as targets. On one side of a microsphere, protons are accelerated by TNSA from the concave inside surface of the sphere. Laser heated electrons that are spreading transversely in the target, as a charge wave, are refocused on the opposite side of the sphere, where they produce a strong but transient charge separation field in an opening located there. Protons passing through the opening at the correct time can thus be post accelerated. We have done two-dimensional PIC simulations that confirm that this process indeed occurs, and that the electrons spread over the sphere as a charge wave. This wave is found to oscillate back and forth over the sphere while decaying in amplitude, forming charge separation fields in the opening at regular intervals. These simulations also show that protons arriving at the correct time, i.e. those protons that have right kinetic energy, are post accelerated. Experimentally we have demonstrated the technical feasibility of preparing and irradiating this type of targets. In addition, the preliminary results show some signatures of post acceleration, although the timing between the electron charge wave and the TNSA protons were far from optimal in this first experiment. Further work with improved relative timing will be needed to fully explore the potential of this new scheme and target geometry.}

\section*{Acknowledgments}
We acknowledge the support from the Swedish Research Council (including e.g. Contract No. 2007-4422 and the Linnaeus grant to the Lund Laser Centre), {the Marie Curie Early Stage Training Site MAXLAS (MEST-CT-2005-020356) within the $6^{th}$ European Framework Programme} , the Knut and Alice Wallenberg Foundation, the Swedish National Infrastructure for Computing (SNIC) and the EPSRC through grant no EP/E048668/1. This research was further partially supported by the European Research Council under Contract No. 204059-QPQV. We also thank D.C. Carroll for the processing of the CR-39 plates.

\end{document}